# A Randomly Expandable Method for Data Layout of RAID Storage Systems

De-zhai YUAN[†‡1,2], Xing-yi PENG[1,2], Ting LIU[1,2], Zhe CUI[1]

*(1 Chengdu Institute of Computer Applications, Chinese Academy of Sciences Chengdu, 610041, China)*
*(²University of Chinese Academy of Sciences Beijing, 100049, China)*
[†]E-mail: yuandezhai12@mails.ucas.ac.cn

**Abstract:** With the increase of huge amounts of data in volume, velocity, and variety, the need for capacity of Redundant Arrays of Inexpensive Disks (RAID) storage systems is dramatically growing. However, the probability of disk failures in RAID storage systems is sharply high with the increase of program/erase cycles, read cycles, and retention time. Furthermore, they are faced with more challenges in fault tolerance, storage efficiency, computational complexity, and expandability. This article presents a novel data layout scheme for RAID storage System using Random Binary Extensive Code (RBEC), which are designed to ensure random expandability, high reliability, and availability of data in RAID storage systems. RBEC is a family of systematic code, in which the generator matrix consists of two submatrices with entries over $GF(2)$, an identity matrix on the top, and another submatrix on the bottom. Compared with the existed approaches, the attractive advantages of our schemes include 1) they are completely implemented based on only simple eXclusive OR (XOR) operations and have systematic code property, 2) they can provide arbitrary fault tolerance, 3) their storage efficiency is quasi-optimal, and 4) data and parity disks of RAID storage systems can be randomly expanded according to practical requirements. Thus, our scheme is particularly suitable for RAID storage systems that need higher reliability, availability, and expandability.

**Key words:** Random Matrix, Data Layout, Fault Tolerance, RAID
**doi:**10.1631/FITEE.1000000      **Document code:**  A      **CLC number:**

## 1 Introduction

Since the storage availability and reliability of Redundant Arrays of Inexpensive Disks (RAID) (Patterson *et al*., 1988) storage systems is seriously degraded with the increase of program/erase cycles, read cycles, and retention time. One of the most urgent challenges is to provide sufficient availability and reliability to prevent data losses and corruption. Consequently, reliable and practical fault-tolerant technologies are required to ensure successful data recovery from several varieties of storage system failures. This kind of technology is divided into two groups: *N*-way mirroring and erasure code. *N*-way mirroring is widely used in actual storage systems, such as GFS (Ghemawat *et al*., 2003), Hadoop (Shvachko *et al*., 2010), and Dynamo (DeCandia *et al*., 2007), whose significance lies in providing additional redundancy to ensure successful recovery, but the storage efficiency is exceedingly low. Due to the low storage spaces utilization efficiency of *N*-way mirroring technique, erasure code is more suitable for storage systems with low redundancy and high fault tolerance compared with *N*-way mirroring technique. Because of the high efficiency and practicability, erasure code has gradually attracted more and more attention from the industry and academe, thus becomes a hot research topic of the field of data storage in recent years.

All kinds of erasure codes have been proposed for RAID storage systems after years of painstaking research and development, especially the widely used Reed-Solomon (RS) (Reed and Solomon, 1960) and parity array codes, including, EVEVODD (Blaum *et al*., 1995) , WEAVER (Hafner, 1995) , FENG codes (Feng *et al*., 2005a, 2005b), and so forth. Each of them has obvious advantages, however, none has become the perfect and the actual standard in the storage systems. On one hand, RS codes is a kind of codes that provide optimal storage efficiency and arbitrarily high fault tolerance, but requires special purpose hardware to enable efficient computation of the Galois field arithmetic on which the codes are based and generally has higher computation costs and complexities. On the other hand, parity array codes are completely based on eXclusive OR (XOR) operations, but have relatively low storage efficiency and also cannot be randomly expanded with the increase of data and parity disks of RAID storage systems according to practical requirements. Although most research on erasures codes for RAID storage systems has been on balancing fault tolerance, storage efficiency, and computation complexity, very few efforts have been made to improve flexible expandability.

---
[‡] Corresponding author
[*] Project supported by Projects supported by the National Natural Science Foundation of China (No. 61501064), the National Basic Research Program of China ("973" Program-2011CB302400), the Sichuan Technology Support Program (Nos. 2014GZ0104 and 2015GZ0088), the Knowledge Innovation Project of Chinese Academy of Sciences(No. KGCX2-YW-105), the West Light Foundation of the Chinese Academy of Sciences (No. Y5C2021100) and the Instrument Research and Development Project of the Chinese Academy of Sciences (No. Y6C2011100).
ORCID: De-zhai YUAN, http://orcid.org/0000-0001-5162-8437

Motivated by the fact that the data disks and parity disks of RAID storage systems cannot be randomly expanded according to practical requirements, we present a randomly expandable method for data layout of RAID storage System, which adopts Random Binary Extensive Code (RBEC) (Chen $at\ al.$, 2016) to encode and decode data of RAID storage systems. Compared with the existed approaches, the attractive advantages of our schemes include 1) they are completely implemented based on only simple XOR operations and have systematic code property which are more efficient than traditional RS codes in terms of computational complexity, 2) they can provide arbitrary fault tolerance, 3) their storage efficiency is quasi-optimal, and 4) data and parity disks of RAID storage systems can be randomly expanded according to practical requirements. Thus, all these advantages make our scheme rather suitable for the RAID storage systems that need high reliability, sufficient availability, and flexible expandability.

This article is organized as follows. In the next section, we first briefly introduce the coding model and some relevant terminologies. Section 3 reviews previous research work related to erasure codes. In section 4, we propose the detailed preliminaries used in the construction of our randomly expandable method. Section 5 describes our proposed scheme. Section 6 provides performance analysis, comparisons, and implementation of the proposed scheme. Finally, conclusions are given in section 7.

## 2 Coding Model and Terminologies

The fundamental concept of erasure codes is to encode the $k$ original data blocks into $n$ encoded data blocks. When $t$ pieces of them are lost, the original data blocks can be reconstructed from the left $n-t$ pieces, such a kind of erasure code is called $(n, k)$ coding model. If $t=n-k$, this also can be called Maximum Distance Separate (MDS) codes, which meet the Singleton Bound and provide optimal storage efficiency (MacWilliams, 1977). And when $k$ out of the $n$ encoded data blocks are identical to the $k$ original data blocks, we call it as systematic code, which is distributed in $n$ nodes of the storage system represented in Fig. 1.

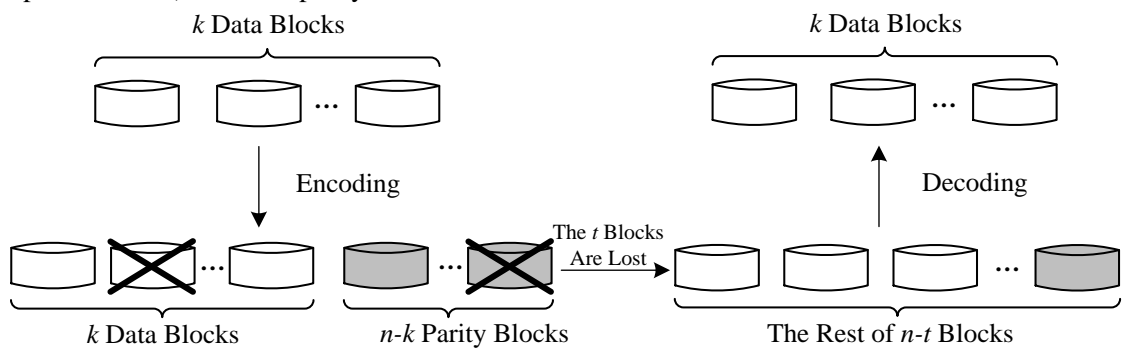

Fig. 1. The ($n, k$) Systematic Code Model

The encoding process between original data blocks and encoded parity blocks can be regarded as a sort of mathematic transform which can be realized by a few classic means such as the Vandermonde matrix and the Cauchy matrix that are widely used in RS codes. In terms of coding theory, the encoding and decoding process of $(n, k)$ coding model can be equivalently expressed as two specific matrices, namely, the generator matrix and the parity-check matrix. The former is used to generate the $n$ encoded data blocks, and the latter reconstruct the $k$ original data blocks. The successful recovery of original data blocks lies in the orthogonality of the generator matrix and the parity-check matrix. Thus, an erasure code can be uniquely specified by the the two matrices in a broad sense, and the key to the successful recovery of original data blocks is to construct such two specific matrices that can meet the orthogonal property.

In order to prevent confusion and better understand the erasure codes and RAID storage systems terms, a number of significant terminologies concerning them will be enumerated in the following. Some of them will also be used throughout this paper for describing and evaluating our scheme referring to related works (Hafner, 2004, 2005, 2006).

- Data: a piece of bits, bytes or blocks that carry user unmodified data.
- Parity: a piece of bits, bytes or blocks that carry redundant information generated from user data applied for data recovery.
- Element: the basic building block of erasure codes usually referring to a unit of data or parity, like bit, byte, sector or larger disk block. In coding theory, this is a bit within a code symbol.
- Stripe: a connected set of data and parity elements that are dependently related by coding. In coding theory, this is a codeword, and its length is usually defined as the number of disks that a stripe stretches over, for example, the $i$-th codeword component is stored on the $i$-th disk.
- Strip: a stripe unit or a maximal set of continuous element in a stripe stored on the same disk. In coding theory, this is a code symbol, and its width is customarily defined as the number of elements consisted in a strip.
- Array: a collection of disks on which one or more stripes are implemented. Each codeword may have a different logical mapping of strip to disk for reasons such as load levelling.
- Stack: a collection of stripes in an array that are related by a maximal set of permutations of logical mappings of strip number to disk. Maximal here means that the losses of any two (or one) physical disks covers all combinations of losses of two (or one) logical strips for the purpose of uniformizing strip failure scenarios under any disk failure case.
- Systematic Code: the codeword is divided into two parts, namely, the data part and the parity part. The data part won't be modified after encoding which can be directly read by users in the case of no errors.
- Vertical Codes: a kind of erasure code in which a strip contains both data elements and parity elements (e.g., WEAVER Codes (Hafner, 1995) and X-codes (Xu and Bruck, 1999)).
- Horizontal Codes: a kind of erasure code in which a strip contains either data elements or parity elements, but a stripe contains both data elements and parity elements (e.g., EVENODD (Blaum *et al.*, 1995) and STAR (Huang *et al.*, 2008)).
- HoVer Codes (Two-dimensional Codes): most of successive strips contain both data elements and parity elements, and the rest of them only hold parity elements. Furthermore, N-dimensional Codes are also included (e.g., HoVer Codes (Hafner, 2006) and GRID (Li *et al.*, 2009)).
- Storage Efficiency: the proportion of a stripe which contains data elements known as the number of data elements divided by the total number of elements including data and parity.
- Fault Tolerance: the maximum number of lost strips which can be reconstructed by erasure codes accurately.
- Complexity: The computational costs of encoding, decoding and updating.

To conceptualize that visually, Fig. 2 represents the data layout of elements, strips, stripes, stacks, and arrays in the typical horizontal codes of RAID storage systems (Hafner, 2004). Apparently, the hierarchical structure of them can be clear at a glance from the figure below.

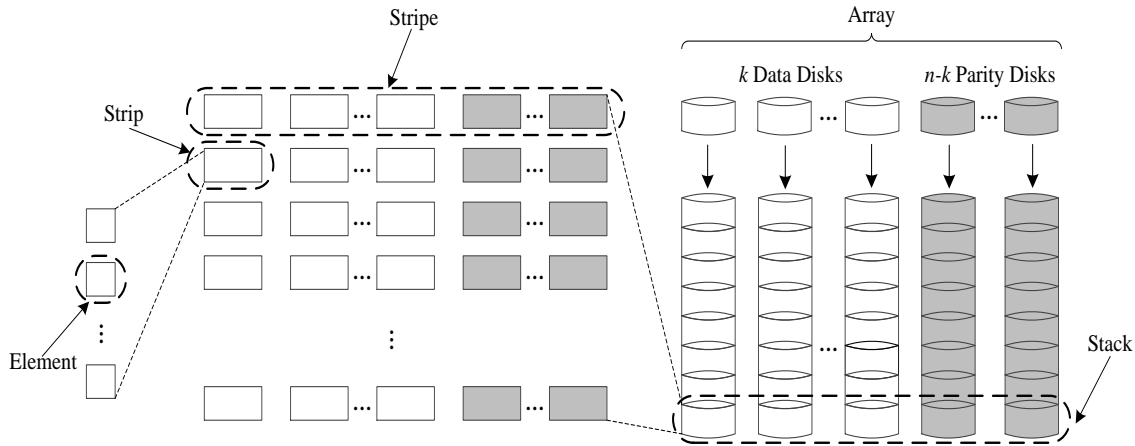

**Fig. 2. The Data Layout of Horizontal Codes in RAID storage systems**

## 3 Related Work

There are many researches on erasures codes for RAID storage systems which has been put forward balancing fault tolerance, storage efficiency, and computation complexity before we represent our scheme. In this paper, we divide the existed erasure codes into three fundamental categories and cite some instances to illustrate them picturesquely referring to Plank and Huang (Plank and Huang, 2013).

Reed-Solomon (RS) Codes: An ancient but often widely used code originated from Reed and Solomon which is based on Vandermonde matrix. RS codes (Reed and Solomon, 1960) are a family of MDS codes that provide optimal storage efficiency and arbitrarily high fault tolerance. However, RS codes are based on Galois field arithmetic over $GF(2^w)$, so they require special purpose hardware to enable efficient computation of the Galois field arithmetic on which the codes are based and generally has higher computation costs and complexities. After constant improvement and innovation, Cauchy_RS Codes (Roth et al., 1989) are represented through Cauchy matrix using XOR-based operations in the form of $GF(2^w)$ by $w \times w$ matrix over $GF(2)$ instead of complicated Galois field arithmetic over $GF(2^w)$. Derived from RS Codes, FENG codes (Feng et al., 2005a, 2005b) are also proposed including Reed-Solomon-Like Code and Rabin-Like Code. For the sake of reducing the complexity of expensive Galois field arithmetic operation, the cyclotomic fast Fourier transform algorithmis also presented in the implementation of RAID based on RS codes, which is much lower than the existed MDS array codes (Trifonov, 2015).

Parity Array Codes: There are at least three types of parity array codes: horizontal codes, for instance, Row Diagonal Parity (RDP) (Corbett et al., 2004), EVENODD (Blaum *et al.*, 1995) and generalized X-Code (Luo *et al.*, 2012); vertical codes, such as, WEAVER (Hafner, 1995), C-Code (Li *et al.*, 2011), X-Code (Xu and Bruck, 1999) and P-Code (Jin *et al.*, 2009); 2-dimensional (or higher N-dimensional) horizontal and vertical code, for example, HoVer codes (Hafner, 2006) and GRID codes (Li *et al.*, 2009). A general character of parity array codes is that they are completely implemented based on only simple XOR operations. This is more efficient than traditional RS codes using complicated Galois field operations for encoding and decoding processes in terms of computational complexity (Li et al., 2011). Nevertheless, so many parity array codes as we may have, none is optimal in that each of them has inherent disadvantages and specific scope of application. Obviously, parity array codes have relatively low storage efficiency and also cannot be randomly expanded with the increase of data and parity disks of RAID storage systems according to practical requirements.

New Codes: In addition to the above mentioned techniques for constructing codes, more and more innovative approaches are presented recently, for example, Low Density Parity Codes (LDPC) (Gallager, 1962), CRC-Detect-First-LDPC (CDF-LDPC) (Qi *et al.*, 2017), Regenerating codes (Dimakis *et al.*, 2010), Sector Disk (SD) codes

(Plank and Blaum, 2014), STAIR codes (Li *et al.*, 2014), HACFS codes (Xia *et al.*, 2015) and Random RAID (Teng *et al.*, 2017). LDPC are completely XOR-based linear codes defined by bipartite graphs with data elements on the left and parity elements on the right, such as Tornado codes (Luby *et al.*, 1997), LT codes (Luby, 2002) and its improvement Raptor codes (Shokrollahi, 2006). CDF-LDPC algorithm is a new error correction method for Solid-State Drive (SSD), which combines error detection code (EDC, such as cyclic redundancy code, parity check code) with error correction code (ECC, such as LDPC) to improve the read performance of SSD. Regenerating codes are designed to decrease bandwidth for recovery by increasing more element blocks than before that each storage node holds. SD codes and STAIR codes are invented to tolerate the mixed failure models, combinations of sector and disk failures simultaneously, which are more efficient than the traditional codes solely tolerating failures of whole disks. HACFS is a novel erasure-coded storage system that uses a fast code to optimize for recovery performance and a compact code to reduce the storage overhead rather than using two different erasure codes. Random RAID is a new kind of storage fault-tolerance method with high fault-tolerance and flexible scalability by probabilistic approach.

## 4 Preliminaries

In this section, we will briefly introduce the definition of random matrix and its excellent properties that will be used in RBEC code (Chen *at al.*, 2016). Besides, RBEC code will also be introduced, which are designed to ensure random expandability, high reliability, and availability of data in RAID storage systems.

### 4.1 Definition of the random matrix

***Definition 4.1*** Let $M = (m_{i,j})_{n \times n}$ be a random $n \times n$ matrix over $GF(2)$ whose entries are independently and identically distributed, which is defined by

$$\Pr(m_{i,j} = r) = \begin{cases} 1-p, & r=0 \\ p, & r=1, \end{cases} \quad (1)$$

where $p$ denotes the probability of entry being 1. For simplicity, let us suppose that $p = \frac{1}{2}$ and $\Pr(m_{i,j} = 0) = \Pr(m_{i,j} = 1) = \frac{1}{2}$, such that all of the matrix elements are equiprobable, homogeneous, and random. The elaborate generating process is described by the following Algorithm 1.

| **Algorithm 1.** Construction of {0, 1} Random Matrix |
|---|
| 1: **Input**: Size of random matrix; |
| 2: **Output**: The generated $n \times n$ random matrix composed of {0, 1} |
| 3: **repeat** |
| 4:   **for** *i* from 1 to *n* step by 1 do |
| 5:     **for** *j* from 1 to *n* step by 1 do |
| 6:       Generate a random floating number between 0 and 1; |
| 7:       $Rnd_{i,j} = Rand( ) / Double(RAND\_MAX)$; |
| 8:       **if** $(0 \leq Rnd_{i,j} \leq 0.5)$ |
| 9:         $m_{i,j} = 0$; |
| 10:      **else** |
| 11:        $m_{i,j} = 1$; |
| 12:      **endif** |
| 13:    **end** |
| 14:  **end** |
| 15: **until** $i = n$ and $j = n$; |

### 4.2 The properties of random matrix

From previous works (Cooper, 2000), we can easily derive the probability of the generated random matrix being nonsingular, then

***Lemma 4.1***

$$\Pr(Rank(M_{n \times n}) = n) = \prod_{i=1}^{n}(1 - \frac{1}{2^i}). \quad (2)$$

***Proof.*** Let $M_{n \times n} = (\eta_1, \eta_2, \ldots, \eta_n)$ be a random matrix consisting *n* columns and let $\eta_i$ be the *i*-th column, where $1 \leq i \leq 1$. $Rank(M_{n \times n}) = n \Leftrightarrow$ Each *i*-th column cannot be linearly combined by the first *i*-1 columns, denoted by

$$L(i) = 1 - \frac{2^{i-1}}{2^n}, \quad 1 \leq i \leq n \quad (3)$$

Where $2^n$ is the totality of the *i*-th column constitution over $GF(2)$, $2^{i-1}$ is the first *i*-1 columns combinations, $\frac{2^{i-1}}{2^n}$ means the probability of the *i*-th column being linearly combined by the first *i*-1 columns. So,

$$\Pr(Rank(M_{n\times n}) = n) = \prod_{i=1}^{n} L(i)$$
$$= \prod_{i=1}^{n}(1 - \frac{2^{i-1}}{2^n})$$
$$= (1 - \frac{2^0}{2^n})(1 - \frac{2^1}{2^n}),\ldots,(1 - \frac{2^{n-1}}{2^n}) \quad (4)$$
$$= (1 - \frac{1}{2^n})(1 - \frac{1}{2^{n-1}}),\ldots,(1 - \frac{1}{2})$$
$$= \prod_{i=1}^{n}(1 - \frac{1}{2^i})$$
$$= S(n,n).$$

where $\prod_{i=1}^{n}(1 - \frac{1}{2^i})$ can be simply expressed in terms of $S(n,n)$ which denotes the probability of the generated random matrix $M_{n\times n}$ being nonsingular.

***Lemma 4.2*** $S(n,n)$ is a monotone decreasing function.

***Proof.*** Let $1 \leq x_1 < x_2 \leq n$, so

$$\frac{S(x_2, x_2)}{S(x_1, x_1)} = \frac{\prod_{i=1}^{x_2}(1 - \frac{1}{2^{x_2}})}{\prod_{i=1}^{x_1}(1 - \frac{1}{2^{x_1}})}$$
$$= (1 - \frac{1}{2^{x_1+1}})(1 - \frac{1}{2^{x_1+2}}),\ldots,(1 - \frac{1}{2^{x_2}}) \quad (5)$$
$$< 1.$$

Thus, $S(n,n)$ is a monotone decreasing function.

The explicit value of $S(n,n)$ hasn't been solved by the universal scientists so far. But we find that the function tends to a constant 0.28879 when $n \geq 10$ through computational simulation. The tendency of $S(n,n)$ is revealed by Fig. 3 in which the *x*-axis represents *n*, and the *y*-axis refers to the approximate tendency of $S(n,n)$.

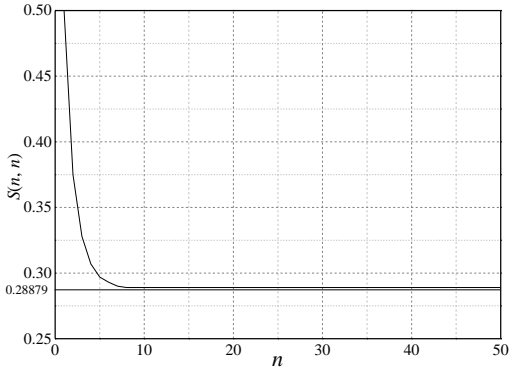

**Fig. 3. The Tendency of $S(n, n)$**

On the basis of the construction of random $n \times n$ matrix and its probability of being nonsingular, the random $(n+k) \times n$ matrix, called high matrix $G_{(n+k)\times n}$, can be easily inferred and the probability of $G_{(n+k)\times n}$ being full column rank is also defined by

***Lemma 4.3***

$$\Pr(Rank(G_{(n+k)\times n}) = n) = \prod_{i=k+1}^{n+k}(1 - \frac{1}{2^i}) \quad (6)$$

***Proof.***

$$\Pr(Rank(G_{(n+k)\times n}) = n) = \prod_{i=0}^{n-1}(1 - \frac{2^i}{2^{n+k}})$$
$$= (1 - \frac{2^0}{2^{n+k}})(1 - \frac{2^1}{2^{n+k}}),\ldots,(1 - \frac{2^{n-1}}{2^{n+k}})$$
$$= (1 - \frac{1}{2^{k+1}})(1 - \frac{1}{2^{k+2}}),\ldots,(1 - \frac{1}{2^{n+k}})$$
$$= \prod_{i=k+1}^{n+k}(1 - \frac{1}{2^i})$$
$$= S(n+k, n).$$
(7)

From $\prod_{i=k+1}^{n+k}(1 - \frac{1}{2^i})$ we can easily come to the conclusion that *n* have little effects on the trend of $S(n+k, k)$ with the increase of *k*. The value of $S(n+k, k)$ is extremely close to 1 when $k \geq 10$, so the tendency of $S(n+k, k)$ is clear as Fig. 4 illustrated.

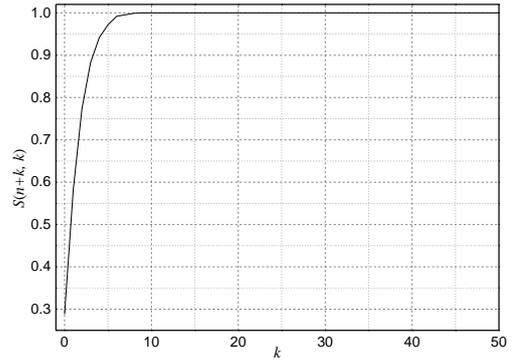

**Fig. 4. The Tendency of S(*n*+*k*, n)**

The above mentioned experiments show that the precision accuracy of $S(n+k, k)$ is exceedingly high when $k \geq 10$.

### 4.3 RBEC

RBEC is a family of systematic code, which can be represented by the generator matrix consisting of two submatrices with entries over $GF(2)$, an identity matrix on the top, and another random submatrix on the bottom. The successful encoding and decoding of RBEC codes primarily lies in the high probability of being full column rank properties of random matrix.

*RBEC Encoding*: The purpose of RBEC encoding is to generate codewords combined together with original data blocks and encoded parity blocks. The RBEC encoding is an actually more efficient process which is a product of original data blocks and the RBEC generator matrix as shown in Eq. (8). For example, $G \cdot D = C$, where $D$ is original data with $k$ blocks, $G_{n \times k}$ is the RBEC generator matrix, and $C$ is the codeword with $n$ blocks. Let $I_{k \times k}$ be an $k \times k$ identity matrix and $R_{(n-k) \times k}$ be a $(n-k) \times k$ random matrix. Now, we define the RBEC generator matrix as follow:

$$G_{n \times k} = \begin{bmatrix} I_{k \times k} \\ R_{(n-k) \times k} \end{bmatrix} = \begin{bmatrix} 1 & 0 & \cdots & 0 \\ 0 & 1 & \cdots & 0 \\ \vdots & \vdots & \ddots & \vdots \\ 0 & 0 & \cdots & 1 \\ r_{1,1} & r_{1,2} & \cdots & r_{1,k} \\ r_{2,1} & r_{2,2} & \cdots & r_{2,k} \\ \vdots & \vdots & \ddots & \vdots \\ r_{n-k,1} & r_{n-k,2} & \cdots & r_{n-k,k} \end{bmatrix}, \quad (8)$$

where $r_{i,j} \in GF(2)$ for $1 \leq i \leq k$ and $1 \leq j \leq n$.

*RBEC Decoding*: The RBEC decoding process can be briefly summarized as the process of reconstructing the original data $D$ by the parity-check matrix $H_{k \times (n-k)}$ which can be derived from $G_{n \times k}$ and defined as shown in Eq. (9).

$$H_{n \times (n-k)} = \begin{bmatrix} R^T_{k \times (n-k)} \\ I_{(n-k) \times (n-k)} \end{bmatrix} = \begin{bmatrix} r_{1,1} & r_{2,1} & \cdots & r_{n-k,1} \\ r_{1,2} & r_{2,2} & \cdots & r_{n-k,2} \\ \vdots & \vdots & \ddots & \vdots \\ r_{1,k} & r_{2,k} & \cdots & r_{n-k,k} \\ 1 & 0 & \cdots & 0 \\ 0 & 1 & \cdots & 0 \\ \vdots & \vdots & \ddots & \vdots \\ 0 & 0 & \cdots & 1 \end{bmatrix}, \quad (9)$$

where $R^T_{k \times (n-k)}$ is the transpose of $R_{(n-k) \times k}$. Besides, it also can be easily checked that $G^T_{k \times n} \times H_{n \times (n-k)} = 0_{k \times (n-k)}$, where $0_{k \times (n-k)}$ is an $k \times (n-k)$ all-zero matrix. Thus, we have $H^T_{(n-k) \times n} \times C_{n \times 1} = 0_{(n-k) \times 1}$, and the decoding process can be reduced to solving system of equations. For more about the improved RBEC decoding, see (Chen *at al.*, 2016).

## 5 Our Proposed Scheme

In this section, we will present a novel data layout scheme for RAID storage System using RBEC code. A basic data layout scheme is provided for RAID, and the randomly expandable method of data and parity disks of RAID according to practical requirements will also be given.

### 5.1 A basic data layout scheme

Suppose that there exist such a RAID storage systems containing 5 data disks numbered 1 through 5, each disk has 4 disk sectors, and 3 parity disks numbered 1, 2, and 3. We first need to initialize a $40 \times 25$ generator matrix $G_{40 \times 25}$ with an $25 \times 25$ identity matrix on the top and another $15 \times 25$ random submatrix on the bottom based on RBEC code. Then, we regard the original data with 25 blocks $(D_1, D_2, D_3, \ldots, D_{24}, D_{25})$ as the message $D$ that need to be encoded. We will make use of message $D$ to generate codeword $W$ with the generator matrix $G_{40 \times 25}$, defined by $W = G_{40 \times 25} \cdot D$. Since the upper part of $G_{40 \times 25}$ is an $25 \times 25$ identity matrix, the former components of the codeword $W$ are identical to message $D$, so the codeword $W$ can be denoted by $W = (D_1, D_2, D_3, \ldots, D_{24}, D_{25}, P_1, P_2, P_3, \ldots, P_{14}, P_{15})$. Finally, for array reasons, the generated one-dimensional codeword $W$ itself will be arranged into $5 \times 8$ two-dimensional array placed in data and parity disks of RAID storage system as Fig. 5 illustrated.

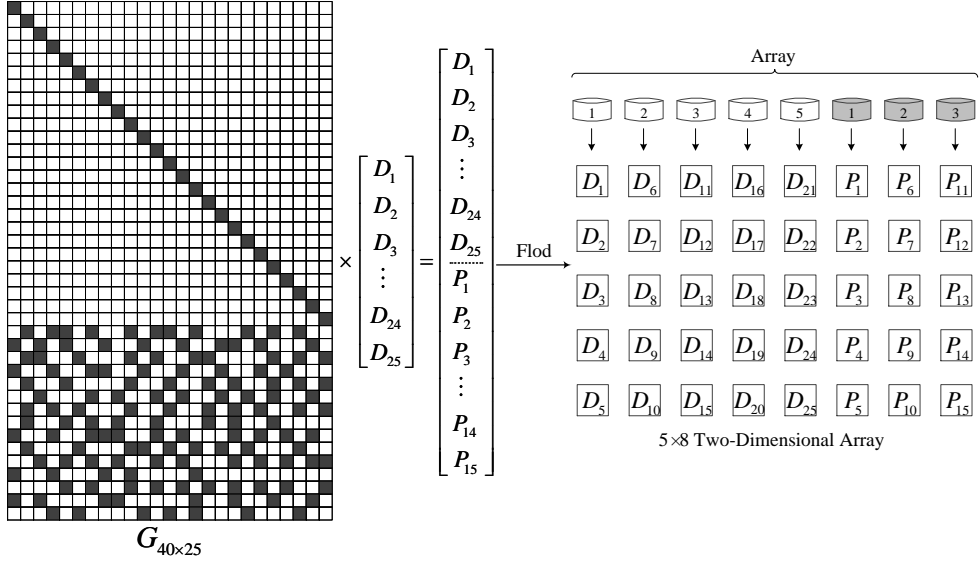

**Fig. 5. The Two-Dimensional Array Data Layout**

In practical applications, the data and parity disks of RAID storage systems can be randomly expanded according to practical requirements. Adding disks can be regarded as the expansion of RAID storage systems. Meanwhile, removing disks also can be regarded as the disk failures in RAID storage systems.

### 5.2 The random expansion of data disks

In this case, the number of data disks will be dynamically adjusted to meet practical requirement with the growth of data volume. Then, we assume that only one data disk will be added or removed in our example. The diagram in Fig. 6 shows their general data/parity layout. We have two cases:

*Removing*: Once one data disk is removed, the corresponding generator matrix will be dynamically adjusted, also and the data and parity disks need to be updated accordingly. Assume that in this example the data disk numbered 2 is removed. Then, the 5 rows numbered 6 through 10, and 5 columns numbered 6 through 10 in generator matrix $G_{40\times25}$ will be removed. The original generator matrix $G_{40\times25}$ will be converted into a newly generated matrix $\bar{G}_{35\times20}$ with an $20\times20$ identity matrix on the top and another $15\times20$ random submatrix on the bottom. Furthermore, the original data with 25 blocks $(D_1, D_2, D_3, \ldots, D_{24}, D_{25})$ as the message $D$ needs to be cut into $\bar{D}$ with 20 blocks again. Meanwhile, message $\bar{D}$ needs to be stored in the rest of 4 data disks numbered 1, 3, 4, and 5. Finally, the parity disks numbered 1, 2, and 3 need to be updated accordingly to $\bar{D}$ and $\bar{G}_{35\times20}$, defined by $\bar{W} = G_{35\times20} \cdot \bar{D}$.

*Adding*: In contrast with the removing case, once one data disk is added, the corresponding generator matrix will be dynamically adjusted, also and the data and parity disks need to be updated accordingly. Assume that in this example the data disk numbered 2 is added. Then, the 5 rows numbered 6 through 10, and 5 columns numbered 6 through 10 in generator matrix $\bar{G}_{35\times20}$ will be added. The original generator matrix $\bar{G}_{35\times20}$ will be converted into a newly generated matrix $G_{40\times25}$ with an $25\times25$ identity matrix on the top and another $15\times25$ random submatrix on the bottom. Furthermore, the original data with 20 blocks $(D_1, D_2, D_3, \ldots, D_{19}, D_{20})$ as the message $\bar{D}$ needs to be cut into $D$ with 25 blocks again. Meanwhile, message $D$ needs to be stored in the 5 data disks numbered 1 through 5. Finally, the parity disks numbered 1, 2, and 3 need to be updated accordingly to $D$ and $G_{40\times25}$, defined by $W = G_{40\times25} \cdot D$.

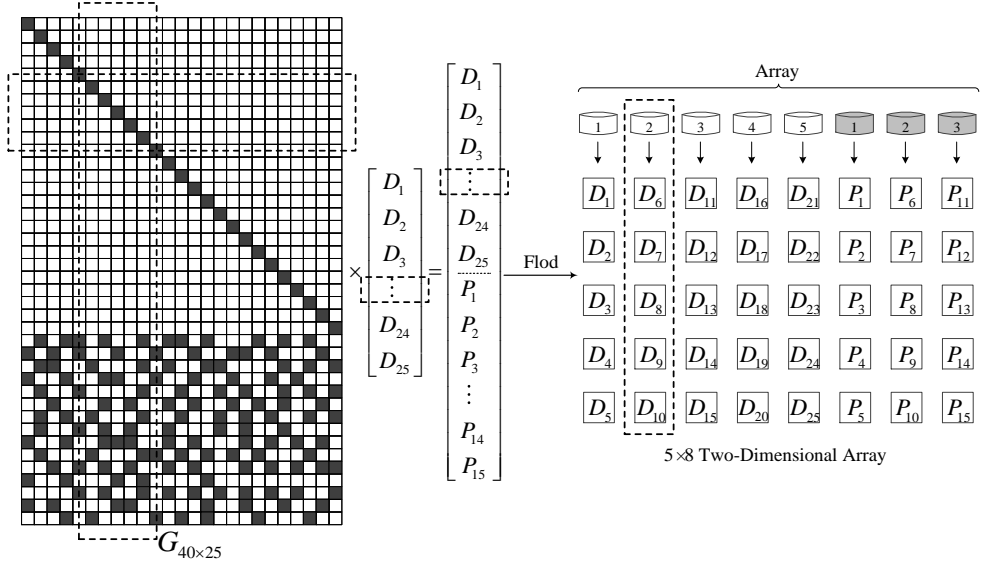

**Fig. 6. The Expansion of Data Disks**

### 5.3 The random expansion of parity disks

This case is similar to the random expansion of data disks, the number of parity disks will be dynamically adjusted to provide reliability for customer data in the presence of RAID storage systems. Then, we assume that only one parity disk will be added or removed in our example. The diagram in Fig. 7 shows their general data/parity layout. We also have two cases:

*Removing*: Once one parity disk is removed, the corresponding generator matrix will be dynamically adjusted, simply but the data and parity disks don't need to be updated accordingly. Assume that in this example the parity disk numbered 3 is removed. Then, the 5 rows numbered 36 through 40 in generator matrix $G_{40 \times 25}$ will be removed. The original generator matrix $G_{40 \times 25}$ will be converted into a newly generated matrix $\bar{G}_{35 \times 25}$ with an $25 \times 25$ identity matrix on the top and another $10 \times 25$ random submatrix on the bottom.

*Adding*: In contrast with the removing case, once one parity disk is added, the corresponding generator matrix will be dynamically adjusted, also and only the parity disks need to be updated accordingly. Assume that in this example the parity disk numbered 3 is added. Then, the 5 rows numbered 36 through 40 in generator matrix $G_{35 \times 25}$ will be added. The original generator matrix $G_{35 \times 25}$ will be converted into a newly generated matrix $\bar{G}_{40 \times 25}$ with an $25 \times 25$ identity matrix on the top and another $15 \times 25$ random submatrix on the bottom. Meanwhile, the data disks numbered 1 through 5 don't need to be updated. The parity disks are computed independently, however, only the parity disks numbered 3 need to be updated accordingly to $D$ and $\bar{G}_{40 \times 25}$, defined by $\bar{W} = \bar{G}_{40 \times 25} \cdot D$.

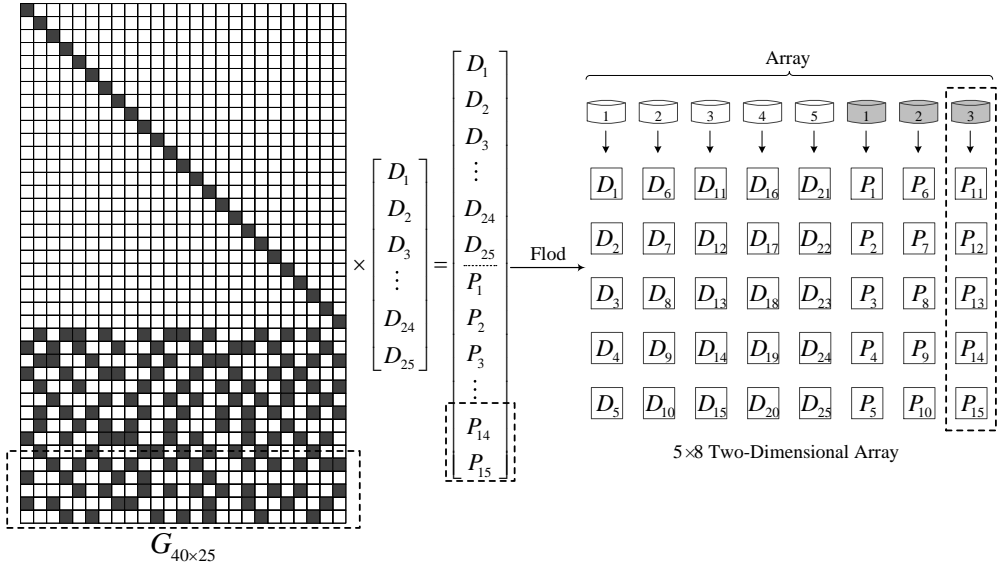

**Fig. 7. The Expansion of Parity Disks**

## 6 Performance and Implementation

In this section, we will summarize some primary features and performances of our scheme, and then compare them with some other existed codes in terms of fault tolerance, storage efficiency, computational complexity, and expandability. Besides, the implementation of our scheme also will be included.

### 6.1 Fault tolerance

From Section 4.1 and (Chen *at al*., 2016, Teng *at al*., 2017,), we can see that for our scheme, its fault tolerance is close to $t = n - k - 10$, where $n$ is the number of data and parity disks, and $k$ is the number of data disks. This further shows that its fault tolerance can be arbitrarily adjusted according to practical requirements dynamically. Especially in the age of big data, our scheme is very suitable for large-scale RAID storage systems in which the possibility of concurrent disk failures, together with multiple unrecoverable sector errors, is very remarkable.

### 6.2 Storage efficiency

From the preceding discussions, we can see that our scheme can provide quasi-optimal storage efficiency, and their storage efficiency can reach up to $e = \dfrac{k}{n} = \dfrac{k}{k+t+10}$, where $k$ is the number of data disks, $t$ is the fault tolerance. It is to be noted that the storage efficiency of our scheme increases with the data disks size and can increase to a very high level, however, only 10 more redundant parity disks need to be provided. It is clear that our scheme with higher fault tolerance always has lower storage efficiency than that with lower fault tolerance. This shows a trade-off between fault tolerance and storage efficiency.

### 6.3 Computational complexity

Our scheme is completely based on XOR operations over Galois field $GF(2)$, and don't need special purpose hardware to enable efficient computation of encoding and decoding over the complex Galois field $GF(2^w)$. The computational complexity of encoding and decoding directly depends on number of 1s in generator matrix $G_{n \times k}$ and parity-check matrix $H_{k \times (n-k)}$. Thus, we can easily deduce that our scheme has an encoding complexity of $O(nk)$ and a decoding complexity of $O(k^3)$.

### 6.4 Expandability

The main difference between other existed approaches is that data and parity disks of RAID storage systems can be randomly expanded according to practical requirements. Furthermore, the original data and parity disks don't need to be completely updated when adding or removing some parity disks. With this capability, we can insert a new disk hot plug into an available slot while the RAID storage system is running.

### 6.5 Comparisons

In this subsection, we will compare our scheme with other existed codes. Some of them are widely used in storage systems and communication fields.

*RS*: Both RS codes and our scheme can provide arbitrarily high fault tolerance and can be randomly expanded according to practical requirements. However, RS codes are based on Galois field arithmetic over $GF(2^w)$ which requires special purpose hardware to enable efficient computation of the Galois field arithmetic on which the codes are based and generally has higher computation costs and complexities. Furthermore, our scheme is completely implemented based on simpler XOR operations instead of complicated Galois field arithmetic. Therefore, our scheme can have much better performance and easier implementation than RS codes.

*Parity Array Codes*: Both parity array codes and our scheme are completely based on XOR operations, and have relatively high storage efficiency. But parity array codes cannot be randomly expanded with the increase of data and parity disks of RAID storage systems according to practical requirements. In our scheme, however, data data and parity disks of RAID storage systems can be randomly expanded. Thus, all these advantages make our scheme rather suitable for the RAID storage systems that need flexible expandability.

*New Codes*: Compared with the newly invented codes, they are designed for some special cases in storage systems, and their structures are too irregular to implement efficiently which are not well suited to RAID storage systems. However, our scheme has very regular structures and thus can be more easily implemented in storage systems, ensuring easy implementation.

**Table 1. Comparison of features with other schemes**

| Schemes | Fault Tolerance | Storage Efficiency | Computational Complexity | Expandability |
|---|---|---|---|---|
| RS | Arbitrary | Optimal | Galois field | Yes |
| EVENODD | 2 | Optimal | XOR | No |
| X-Code | 2 | Optimal | XOR | No |
| HoVer Code | 4 | Quasi-optimal | XOR | No |
| GRID | Up to 15 or even higher | Non-optimal | XOR | No |
| LDPC | Arbitrary | Quasi-optimal | XOR | No |
| Our Scheme | Arbitrary | Quasi-optimal | XOR | Yes |

Tbale 1 compares our scheme with some other schemes in terms of fault tolerance, storage efficiency, computational complexity, and expandability, it is worth noting that our scheme is relatively suitable for RAID storage systems that need flexible expandability, arbitrary fault tolerance, and simple computations.

### 6.6 Implementation

The implementation of our scheme's encoding and decoding is straightforward, which simply follows the procedure described in section 5. Experiments are conducted to compare the encoding and decoding complexity of our proposed scheme with some widely used schemes. The particular schemes compared in the experiments are RS code (Reed and Solomon, 1960) and Cauchy_RS code (Roth *et al.*, 1989). Both of them are expandable codes can provide arbitrarily high fault tolerance. To make the comparison as fair as possible, we use the widely adopted and highly optimized software-based erasure coding implementation, i.e., the Jerasure 2.0 package.

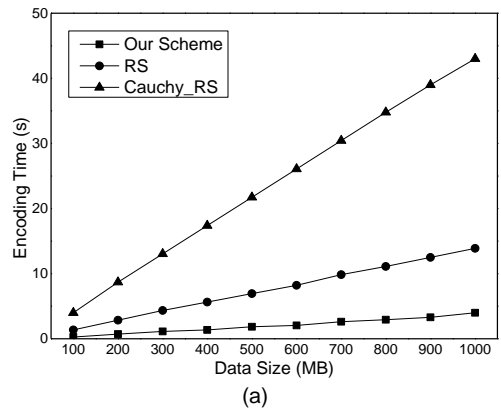

(a)

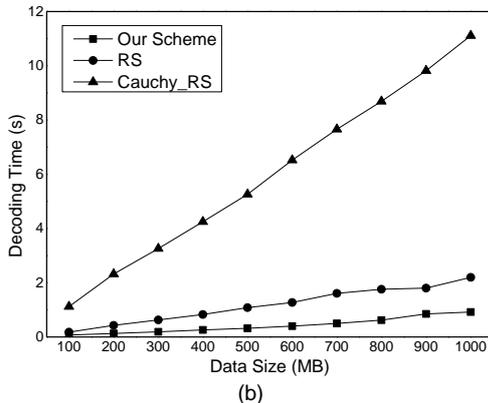

(b)

**Fig. 7. The Computation Time for Encoding and Decoding of some existed schemes. (a) Encoding time. (b) Decoding time**

Fig. 7 shows the computation time of data encoding and decoding using different schemes under various data sizes. The experiments are carried out on a Core i3 2.10-GHz machine with a 4-Gbyte memory running Linux Ubuntu 16.04. It is clear that the proposed scheme still outperforms the widely used expandable schemes by a significant margin.

## 7 Conclusions

In this article, we have presented a novel data layout scheme using RBEC code, which are designed to ensure random expandability, high reliability, and availability of data in RAID storage systems. Compared with the existed approaches, our scheme has the attractive advantages: 1) they are completely implemented based on only simple XOR operations and have systematic code property, ensuring easy implementations 2) they can provide arbitrary fault tolerance only if providing 10 more redundant parity disks, 3) their storage efficiency is quasi-optimal with the growth of data disks of RAID storage systems, and 4) data and parity disks of RAID storage systems can be randomly expanded according to practical requirements. Our scheme provides the designers of RAID storage systems with good tradeoffs between fault tolerance and storage efficiency with the continuous increase of RAID storage system disks. All these advantages make our scheme particularly suitable for large-scale RAID storage systems that need higher reliability, availability, and expandability.

**Acknowledgements**

We are very grateful to the anonymous referees for their insightful comments and suggestions on a previous version of this paper.

**References**

Blaum M, Brady J, Bruck J, *et al.*, 1995. EVENODD: An efficient scheme for tolerating double disk failures in RAID architectures. *IEEE Transactions on computers*, **44**(2):192-202. [doi:10.1109/12.364531]

Chen L, Zhang J Z, Teng P G, *et al.*, 2016. Random Binary Extensive Code (RBEC): An Efficient Code for Distributed Storage System. *Chinese Journal of Computers*.

Cooper C, 2000. On the rank of random matrices. *Random Structures & Algorithms*, **16**(2):209-232. [doi:10.15961/j.jsuese.201600492]

Corbett P, English B, Goel A, *et al.*, 2004. Row-diagonal parity for double disk failure correction. *Proc. of the 3rd USENIX Conference on File and Storage Technologies*, p. 1-14.

DeCandia G, Hastorun D, Jampani M, *et al.*, 2007. Dynamo: amazon's highly available key-value store. *ACM SIGOPS operating systems review*, **41**(6):205-220. [doi:10.1145/1323293.1294281]

Dimakis A G, Godfrey P B, Wu Y, *et al.*, 2010. Network coding for distributed storage systems. *IEEE Transactions on Information Theory*, **56**(9):4539-4551. [doi:10.1109/TIT.2010.2054295]

Feng G L, Deng R H, Bao F, *et al.*,2005a. New Efficient MDS Array Codes for RAID Part I: Reed-Solomon-Like Codes for Tolerating Three Disk Failures. *IEEE Transactions on Computers*, **54**,(9):1071-1080. [doi:10.1109/TC.2005.150]

Feng G L, Deng R H, Bao F, *et al.*, 2005b. New Efficient MDS Array Codes for RAID Part II: Rabin-like codes for tolerating multiple Disk Failures. *IEEE Transactions on Computers*,**54**(12):1473-1483. [doi:10.1109/TC.2005.200]

Gallager R, 1962. Low-density parity-check codes. *IRE Transactions on information theory*, 8(1): 21-28.

Ghemawat S, Gobioff H, and Leung S T, 2003. The Google file system. *Proc. of the 19th ACM symp. on Operating systems principles*, Bolton Landing, New York, USA.

Hafner J L, 2005. WEAVER codes: highly fault tolerant erasure codes for storage systems. *Proc. of the 4th USENIX Conf. on File and Storage Technologies*, p. 211-224, USENIX Association.

Hafner J L, Deenadhayalan V, Kanungo T, *et al.*, 2004. Performance metrics for erasure codes in storage systems. Tech. rep. RJ 10321 (A0408-003). IBM Research Division, Almaden Research Center. August.

Hafner J L, 2006. Hover erasure codes for disk arrays. *Proc. of the Annual IEEE/IFIP Int. Conference on Dependable Systems and Networks*, IEEE Computer Society, p. 217–226. [doi:10.1109/DSN.2006.40]

Hafner J L, 2005. Weaver codes: Highly fault tolerant erasure codes for storage systems. *Proc. of the 4th USENIX Conf. on File and Storage Technologies*, USENIX Association, p. 211–224.

Huang C, Xu L H, 2008. STAR: An efficient coding scheme for correcting triple storage node failures. *IEEE Transactions on Computers*, **57**(7):889-901.[doi: 10.1109/TC.2007.70830]

Jin C, Jiang H, Feng D, et al, 2009. P-Code: A new RAID-6 code with optimal properties. *Proc. of the 23rd int. conf. on Super computing*, p. 360-369, ACM. [doi:10.1145/1542275.1542326]

Li M Q, Shu J W, Zheng W M, 2009 . GRID codes: Strip-based erasure codes with high fault tolerance for storage systems. *ACM Transactions on Storage*, **4**(4):15:1-15:22. [doi:10.1145/1480439.1480444]

Li M Q, Shu J W, 2011. C-Codes: Cyclic Lowest-Density MDS Array Codes Constructed Using Starters for RAID 6. Technical report, Online at: http://arxiv.org/abs/1104.2547.

Li M Q, Lee P P C, 2014 . STAIR codes: A general family of erasure codes for tolerating device and sector failures," *ACM Transactions on Storage*, **10**(4):14:1-14:30.

Luby M G, Mitzenmacher M, Shokrollahi M A, et al, 1997. Practical loss-resilient codes. *Proc. of the twenty-ninth annual ACM symp. on Theory of computing*, p. 150-159, ACM. [doi:10.1145/258533.258573]

Luby M, 2002. LT codes. *Proc. Of the 43rd Annual IEEE Symp. on Foundations of Computer Science*, p. 271–280, Vancouver, BC, Canada, November. [doi: 10.1109/SFCS.2002.1181950]

Luo X H, Shu J W, 2012. Generalized X-Code: An efficient RAID-6 code for arbitrary size of disk array. *ACM Transactions on Storage*, **8**(3)3:10:1-10:16. [doi:10.1145/2339118.2339121]

MacWilliams F J, Sloane N J A, 1977. *The theory of error-correcting codes*, Elsevier. [doi:10.1137/1022103]

Patterson D A, Gibson G, Katz R H, 1988. A case for redundant arrays of inexpensive disks (RAID). *Proc. of the 1988 ACM SIGMOD int. conf. on Management of data*, p.109-116, ACM New York,. [doi:10.1145/971701.50214]

Plank J S, Huang C, 2013. Tutorial: Erasure Coding for Storage Systems. *11th USENIX Conference on File and Storage Technologies*, San Jose, CA, February, 12.

Plank J S, Blaum M, 2014. Sector-disk (SD) erasure codes for mixed failure modes in RAID systems. *ACM Transactions on Storage*, **10**(1)1:4:1-4:17,. [doi:10.1145/2560013]

Qi S, Feng D, Su N, et al, 2017. CDF-LDPC: A New Error Correction Method for SSD to Improve the Read Performance. *ACM Transactions on Storage,* **13**(1): 7:1-7:22, [doi:10.1145/3017430]

Reed I S, Solomon G, 1960. Polynomial codes over certain finite fields," Journal of the society for industrial and applied mathematics, **8**(2):300-304. [doi: 10.2307/2098968]

Roth R M, Lempel A, 1989. On MDS codes via Cauchy matrices. *IEEE Transactions on Information Theory*, **35**(6):1314-1319. [doi:10.1109/18.45291]

Shokrollahi A. Raptor codes. *IEEE transactions on information theory*, **52**(6):2551-2567. [doi:10.1109/TIT.2006.874390]

Shvachko K, Kuang H, Radia S, et al, 2010. The hadoop distributed file system. *Proc. of the 2010 IEEE 26th symp. on Mass storage systems and technologies*, p.1-10, IEEE. [doi:10.1109/MSST.2010.5496972]

Teng P G, Zhang J Z, Chen L, et al, 201. Random RAID: A RAID storage scheme with high fault-tolerance and flexibility," Advanced Engineering Sciences. **49**(3): 110–116. [doi: 10.15961/j.jsuese.201600492]

Trifonov P. Low-complexity implementation of RAID based on Reed-Solomon codes. *ACM Transactions on Storage*, **11**(1)1:1-1:25. [doi:10.1145/2700308]

Xu L, Bruck J, 1999. X-code: MDS array codes with optimal encoding. *IEEE Transactions on Information Theory*, 45(1):272-276. [doi:10.1109/18.746809]

Xia M Y, Saxena M, Blaum M, et al, 2015. A Tale of Two Erasure Codes in HDFS. *Proc. of the 13th USENIX Conf. on File and Storage Technologies*, p.213-226, SantanClara, USA. [doi:10.1109/ICDE.2012.28]